\documentclass[twocolumn,showpacs,preprintnumbers,prl]{revtex4}
\usepackage{psfrag,graphicx,epsfig,amsmath,amssymb}

\newcommand{\beq}{\begin{equation}}
\newcommand{\eeq}{\end{equation}}
\newcommand{\bea}{\begin{eqnarray}}
\newcommand{\eea}{\end{eqnarray}}
\newcommand{\Fig}[1]{Fig.\,\ref{#1}}

\newcommand{\Eq}[1]{Eq.\,(\ref{#1})}

\newcommand{\Eqsand}[2]{Eqs.\,(\ref{#1}) and (\ref{#2})}

\newcommand{\as}{\alpha_s}

\newcommand{\mb}{m_b}

\newcommand{\mh}{m_h}

\newcommand{\GeV}{{\rm GeV}}
\newcommand{\TeV}{{\rm TeV}}

\newcommand{\ord}{{\cal O}}
\def\unit{\leavevmode\hbox{\small1\kern-3.6pt\normalsize1}}

\newcommand{\BXdga}{\bar{B} \to X_d \gamma}
\newcommand{\BXsga}{\bar{B} \to X_s \gamma}

\newcommand{\BRga}{{\cal B} (\BXsga)}

\newcommand{\btosgamma}{b \to s \gamma}
\newcommand{\btosgluon}{b \to s g}

\newcommand{\Ztobb}{Z \to b \bar{b}}

\newcommand{\CP}{C\hspace{-0.25mm}P}

\newcommand{\mysigma}{\hspace{0.4mm} \sigma}

\newcommand{\etal}{{\it et al}.}

\begin{document}

\allowdisplaybreaks

\preprint{ZU-TH 4/07; CLNS 07/1992} 

\title{
\boldmath
Bound on minimal universal extra dimensions from $\BXsga$
\unboldmath
}

\author{Ulrich~Haisch$^1$ and Andreas~Weiler$^{2}$} 

\affiliation{
$^1\!\!\!$ Institut f\"ur Theoretische Physik, Universit\"at Z\"urich,
CH-8057 Z\"urich, Switzerland \\
$^2\!\!\!$ Institute for High-Energy Phenomenology Newman Laboratory
for Elementary-Particle Physics, Cornell University Ithaca, NY 14853,
U.S.A. 
}

\date{\today}

\begin{abstract}
\noindent
We reexamine the constraints on universal extra dimensional models
arising from the inclusive radiative $\BXsga$ decay. We take into
account the leading order contributions due to the exchange of
Kaluza-Klein modes as well as the available next-to-next-to-leading
order corrections to the $\BXsga$ branching ratio in the standard
model. For the case of one large flat universal extra dimension, we
obtain a lower bound on the inverse compactification radius $1/R > 600
\, \GeV$ at $95 \%$ confidence level that is independent of the Higgs
mass.
\end{abstract}


\maketitle

The branching ratio of the inclusive radiative $\bar{B}$-meson decay
is known to provide stringent constraints on various non-standard
physics models at the electroweak scale, because it is accurately
measured and its theoretical determination is rather precise. 

The present experimental world average which includes the latest
measurements by CLEO \cite{Chen:2001fj}, Belle
\cite{Koppenburg:2004fz}, and BaBar \cite{Aubert:2006gg} is performed
by the Heavy Flavor Averaging Group \cite{unknown:2006bi} and reads
for a photon energy cut of $E_\gamma > E_0$ with $E_0 = 1.6 \, \GeV$
in the $\bar{B}$-meson rest-frame
\beq \label{eq:WA} 
\BRga_{\rm exp} =
(3.55 \pm 0.24^{+0.09}_{-0.10} \pm 0.03) \times 10^{-4} \, .
\eeq
Here the first error is a combined statistical and systematic one,
while the second and third are systematic uncertainties due to the
extrapolation from $E_0 = (1.8 - 2.0) \, \GeV$ to the reference value
and the subtraction of the $\BXdga$ event fraction, respectively.

After a joint effort \cite{NNLO, Misiak:2006ab, Czakon:2006ss}, the
first theoretical determination of the total $\BXsga$ branching ratio
at next-to-next-to-leading order (NNLO) QCD has been presented
recently in \cite{Misiak:2006ab, Misiak:2006zs}. In
\cite{Becher:2006pu} this fixed-order result has been supplemented
with perturbative cut-related $\ord (\as^2)$ corrections \cite{BN} and
an estimate of enhanced $\Lambda_{\rm QCD}/\mb$ non-local power
corrections using the vacuum insertion approximation
\cite{Lee:2006wn}. For $E_0 = 1.6 \, \GeV$ the result of the improved 
standard model (SM) evaluation is given by \footnote{The small NNLO
  corrections related to the four-loop $\btosgluon$ mixing diagrams
  \cite{Czakon:2006ss} and from the charm quark mass effects of the
  electromagnetic dipole operator contribution \cite{Asatrian:2006rq}
  is not included in the numerical result.}
\beq \label{eq:NNLO}
\BRga_{\rm SM} = (2.98 \pm 0.26) \times 10^{-4} \, , 
\eeq
where the uncertainties from higher-order perturbative effects
$(^{+4}_{-6} \%)$, hadronic power corrections ($\pm 5 \%$), parametric
dependencies ($\pm 4 \%$), and the interpolation in the charm quark
mass ($\pm 3 \%$) have been added in quadrature to obtain the total
error.

Compared with the experimental world average of \Eq{eq:WA}, the new SM
prediction of \Eq{eq:NNLO} is lower by around $1.4
\mysigma$. Potential beyond SM contributions should now be preferably
constructive, while models that lead to a suppression of the
$\btosgamma$ amplitude are more severely constrained than in the past,
where the theoretical determination used to be above the experimental
one.

\begin{figure}[t!]
\begin{center}
\vspace{2mm}
\makebox{
\begin{psfrags}
\newcommand{\psfragtextscale}{1}
\providecommand{\psfragtextscale}{1}
\providecommand{\psfragmathscale}{\psfragtextscale}
\providecommand{\psfragnumericscale}{\psfragtextscale}
\providecommand{\psfragtextstyle}{}
\providecommand{\psfragmathstyle}{}
\providecommand{\psfragnumericstyle}{}

\psfrag{x}[cc][cc][1.1][0]{$1/R~[{\rm TeV}]$}
\psfrag{y}[bc][bc][1.1][0]{${\cal B} (\bar{B} \to X_s \gamma)~[10^{-4}]$}
\makebox{\hspace{-1.1cm} \includegraphics[width=3.5in,height=2.25in]{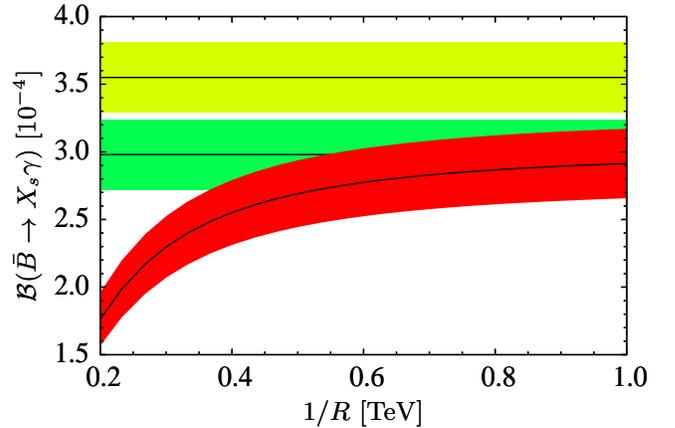}}  
\end{psfrags}
}
\end{center}
\vspace{-4mm}
\caption{\sf $\BRga$ for $E_0 = 1.6 \, \GeV$ as a function of
  $1/R$. The red (dark gray) band corresponds to the LO mUED
  result. The $68 \%$ CL range and central value of the
  experimental/SM result is indicated by the yellow/green
  (light/medium gray) band underlying the straight solid line. See
  text for details.}
\label{fig:ACDLO}
\end{figure}

Among the scenarios of the latter category is the model of Appelquist,
Cheng, and Dobrescu (ACD) \cite{Appelquist:2000nn} as emphasized in
\cite{Agashe:2001xt, Buras:2003mk}. In the ACD framework the SM is
extended from four-dimensional Minkowski space-time to five dimensions
and the extra space dimension is compactified on the orbifold
$S^1/Z_2$ in order to obtain chiral fermions in four dimensions. The
five-dimensional fields can equivalently be described in a
four-dimensional Lagrangian with heavy Kaluza-Klein (KK) states for
every field that lives in the fifth dimension or bulk. In the ACD
model all SM fields are promoted to the bulk. The orbifold
compactification breaks KK number conservation, but preserves
KK-parity. This property implies, that KK states can only be
pair-produced, that their virtual effect comes only from loops, and
causes the lightest KK particle (LKP) to be stable, therefore
providing a viable dark matter (DM) candidate \cite{ueddm} with
promising prospects for direct and indirect detection. See
\cite{Hooper:2007qk} for a recent review on the DM and collider
phenomenology of the ACD model.

The full Lagrangian of the ACD model includes both bulk and boundary
terms. The bulk Lagrangian is determined by the SM parameters after an
appropriate rescaling. The coefficients of the boundary terms,
however, although volume suppressed, are free parameters and will get
renormalized by bulk interactions. Flavor non-universal boundary terms
would lead to unacceptably large flavor-changing-neutral-currents
(FCNCs). In the following we will assume vanishing boundary terms at
the cut-off scale and that the ultraviolet completion does not
introduce additional sources of flavor and $\CP$ violation beyond the
ones already present in the model. These additional assumptions define
the minimal universal extra dimension (mUED) model which belongs to
the class of constrained minimal-flavor-violating \cite{MFV}
scenarios. With this choice contributions from boundary terms are of
higher order \footnote{Boundary terms arise radiatively
  \cite{uedbrane}. They effect the $\btosgamma$ amplitude first at the
  two-loop level. Since we perform a LO analysis of $\BRga$ in the
  mUED model its consistent to neglect these effects.}  and one only
has to consider the bulk Lagrangian in leading order (LO) calculations
in the ACD model.

Since at LO the $b \to s \gamma$ amplitude turns out to be cut-off
independent \cite{Buras:2003mk} the only additional parameter entering
the mUED prediction of $\BRga$ relative to the SM is the inverse of
the compactification radius $1/R$. For a light Higgs mass of $\mh =
115 \, \GeV$ a careful analysis of oblique corrections
\cite{Gogoladze:2006br} gives a lower bound of $1/R > 600 \, \GeV$ at
$90 \%$ confidence level (CL), well above current collider limits of
$1/R \gtrsim 300 \, \GeV$ \cite{acdcollider}. With increasing Higgs
mass the former constraint relaxes significantly leading to $1/R > 300
\, \GeV$ at $90 \%$ CL for $\mh = 500 \, \GeV$
\cite{Gogoladze:2006br}. Other constraints on $1/R$ that derive from
the $\Ztobb$ pseudo observables \cite{Oliver:2002up}, the muon
anomalous magnetic moment $(g - 2)_\mu$ \cite{Appelquist:2001jz}, and
several FCNC processes \cite{Buras:2003mk, Buras:2002ej, acdflavor}
are with $1/R \gtrsim (200-250) \, \GeV$ in general weaker.

Values of $1/R$ as low as $300 \, \GeV$ would also lead to an exciting
phenomenology in the next generation of colliders \cite{Hooper:2007qk,
  acdcollider} and could be of interest in connection with DM searches
\cite{ueddm}. Collider measurements alone do not place an upper bound
on $1/R$. LKPs would overclose the universe for $1/R \gtrsim 1.5 \,
\TeV$ \cite{ueddm}, providing motivation for considering weak-scale KK
particles.

The purpose of this Letter is to point out that combining the present
experimental world average with the improved SM prediction of $\BRga$
forces the compactification scale $1/R$ of the mUED model to lie above
$600 \, \GeV$ if errors are treated as Gaussian. This $95 \%$ CL
exclusion bound is independent of the Higgs mass and therefore
stronger than the constraint that follows from electroweak precision
data. The possibility to derive such a powerful bound has already been
anticipated in \cite{Buras:2003mk}.

\begin{figure}[t!]
\begin{center}
\vspace{-2mm}
\makebox{\scalebox{0.525}{\hspace{-1cm} \includegraphics{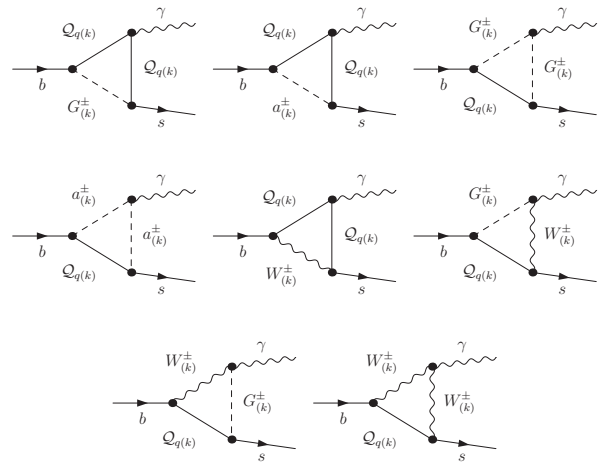}}}
\end{center}
\vspace{-8mm}
\caption{\sf One-loop corrections to the $\btosgamma$ amplitude in the
  mUED model involving infinite towers of the KK modes. Diagrams where
  the $SU(2)$ quark doublets ${\cal Q}_{q (k)}$ are replaced by the
  $SU(2)$ quark singlets ${\cal U}_{q (k)}$ are not shown. See text
  for details.}
\label{fig:ACDdiagrams}
\end{figure}

\begin{figure}
\begin{center}
\makebox{
\begin{psfrags}
\providecommand{\psfragtextscale}{1.2}
\providecommand{\psfragmathscale}{\psfragtextscale}
\providecommand{\psfragnumericscale}{\psfragtextscale}
\providecommand{\psfragtextstyle}{}
\providecommand{\psfragmathstyle}{}
\providecommand{\psfragnumericstyle}{}

\psfrag{x}[cc][cc][1.1][0]{${\cal B} (\bar{B} \to X_s \gamma)_{\rm exp}~[10^{-4}
]$}
\psfrag{y}[bc][bc][1.1][0]{$\Delta {\cal B} (\bar{B} \to X_s \gamma)_{\rm exp}~[
10^{-4}]$}
\makebox{\hspace{-1cm} \includegraphics[width=3.25in,height=3.25in]{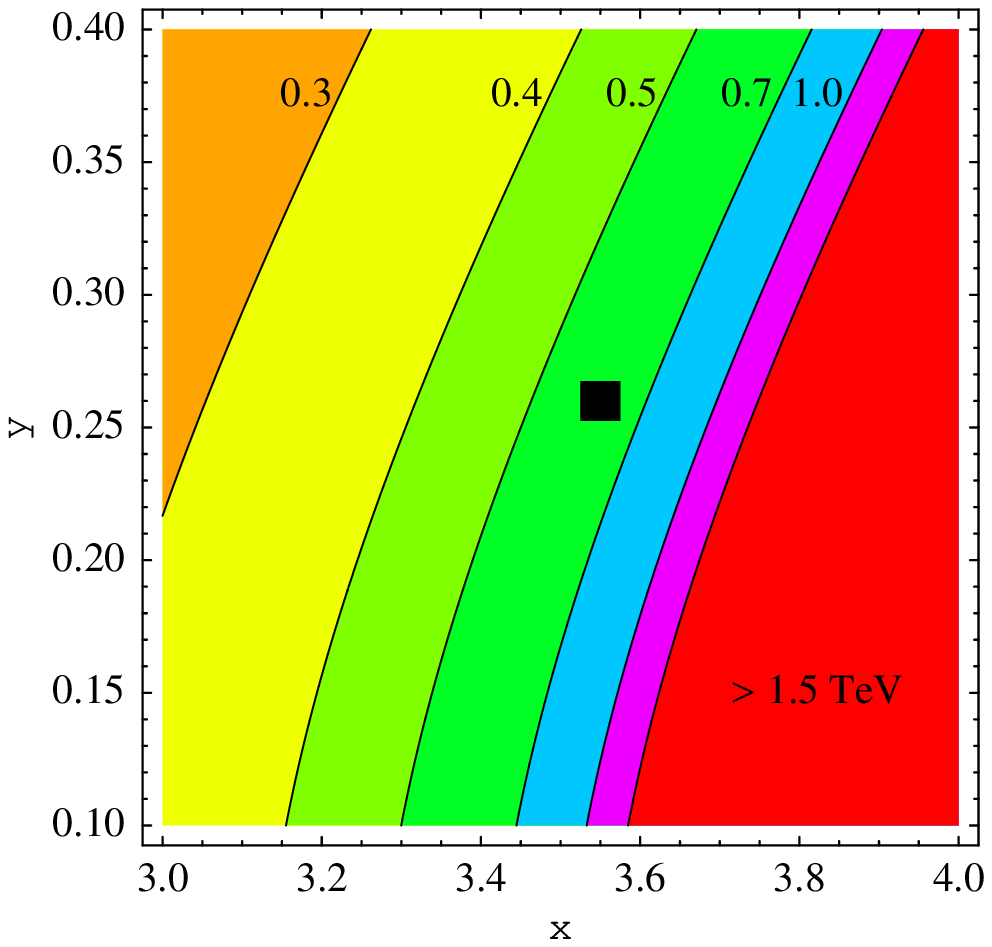}} 
\end{psfrags}
}
\makebox{
\begin{psfrags}
\providecommand{\psfragtextscale}{1.2}
\providecommand{\psfragmathscale}{\psfragtextscale}
\providecommand{\psfragnumericscale}{\psfragtextscale}
\providecommand{\psfragtextstyle}{}
\providecommand{\psfragmathstyle}{}
\providecommand{\psfragnumericstyle}{}

\psfrag{x}[cc][cc][1.1][0]{${\cal B} (\bar{B} \to X_s \gamma)_{\rm SM}~[10^{-4}]
$}
\psfrag{y}[bc][bc][1.1][0]{$\Delta {\cal B} (\bar{B} \to X_s \gamma)_{\rm SM}~[1
0^{-4}]$}
\makebox{\hspace{-1cm} \includegraphics[width=3.25in,height=3.25in]{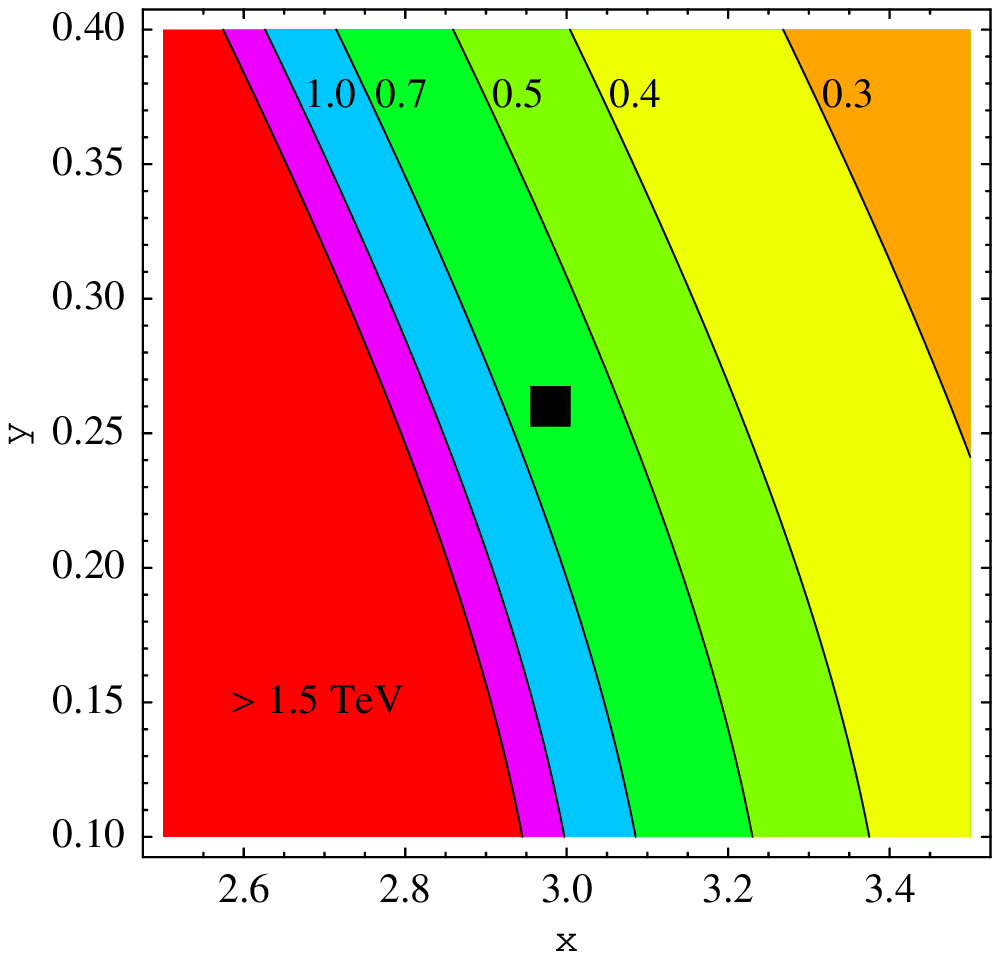}} 
\end{psfrags}
}
\end{center}
\vspace{-6mm}
\caption{\sf The upper/lower panel displays the $95 \%$ CL limits on
  $1/R$ as a function of the experimental/SM central value (horizontal
  axis) and total error (vertical axis). The experimental/SM result
  from \Eq{eq:WA}/\Eq{eq:NNLO} is indicated by the black square. The
  contour lines represent values that lead to the same bound in
  $\TeV$. See text for details.}
\label{fig:ACDbounds}
\end{figure}

The mUED prediction of $\BRga$ for $E_0 = 1.6 \, \GeV$ as a function
of $1/R$ is displayed by the red (dark gray) band in
\Fig{fig:ACDLO}. The yellow (light gray) and green (medium gray) band
in the same figure shows the experimental and SM result as given in
\Eqsand{eq:WA}{eq:NNLO}, respectively. In all three cases, the middle
line is the central value, while the widths of the bands indicate the
uncertainties that one obtains by adding individual errors in
quadrature. The strong suppression of $\BRga$ in the mUED model with
respect to the SM expectation \cite{Agashe:2001xt, Buras:2003mk} and
the slow decoupling of KK modes is clearly seen in \Fig{fig:ACDLO}.

In our numerical analysis, matching of the mUED Wilson coefficients at
the electroweak scale is complete up to the LO \cite{Buras:2003mk},
while terms beyond that order include SM contributions only. The LO
mUED matching correction to the $\btosgamma$ amplitude is found from
the one-loop diagrams that can be seen in \Fig{fig:ACDdiagrams}. The
shown Feynman graphs have been calculated first in
\cite{Buras:2003mk}. They contain apart from the ordinary SM fields,
infinite towers of the KK modes corresponding to the $W$-boson,
$W^\pm_{(k)}$, the pseudo Goldstone boson, $G^\pm_{(k)}$, the $SU (2)$
quark doublets, ${\cal Q}_{q (k)}$, and singlets, ${\cal U}_{q
  (k)}$. Additionally, there appears a charged scalar, $a^\pm_{(k)}$,
which has no counterpart in the SM. The uncertainty related to higher
orders in the mUED model is estimated by varying the matching scale
between $80$ and $320 \, \GeV$. It does not exceed $^{+8}_{-9} \%$ for
$1/R$ in the range of $0.2$ and $1.5 \, \TeV$. Whether this provides a
reliable estimate of the next-to-leading order (NLO) QCD corrections
to $\BRga$ in the mUED model can only be seen by performing a complete
two-loop matching involving KK gluon corrections. Such a calculation
seems worthwhile but it is beyond the scope of this Letter.

Since the experimental result is at present above the SM one and KK
modes in the mUED model interfere destructively with the SM
$\btosgamma$ amplitude, the lower bound on $1/R$ following from
$\BRga$ turns out to be stronger than what one can derive from any
other currently available measurement. If all the uncertainties are
treated as Gaussian and combined in quadrature, the $95 \%$ ($99 \%$)
CL bound amounts to $600$ ($330$) $\GeV$. In contrast to the limit
coming from electroweak precision measurements the latter exclusion is
almost independent of the Higgs mass because genuine electroweak
effects related to Higgs exchange enter $\BRga$ first at the two-loop
level. In the SM these corrections have been calculated
\cite{paolouli} and amount to around $-1.5 \%$ in the branching
ratio. They are included in \Eq{eq:NNLO}. Neglecting the corresponding
two-loop Higgs effects in the mUED model calculation should therefore
have practically no influence on the derived limits.

The upper (lower) contour plot in \Fig{fig:ACDbounds} shows the $95
\%$ CL bound of $1/R$ as a function of the experimental (SM) central
value and error. The current experimental world average and SM
prediction of \Eqsand{eq:WA}{eq:NNLO} are indicated by the black
squares. These plots allow to monitor the effect of future
improvements in both the measurements and the SM prediction. One
should keep in mind, of course, that the derived bounds depend in a
non-negligible way on the treatment of theoretical uncertainties.
Furthermore, the found limits could be weakened by the NLO QCD
matching corrections in the mUED model which remain unknown.

To conclude, we have pointed out that combining the present
experimental with the improved standard model result for the branching
ratio of the inclusive $\BXsga$ decay implies that the inverse
compactification radius of the minimal universal extra dimension model
has to satisfy $1/R > 600 \, \GeV$ at $95 \%$ confidence level if all
the uncertainties are treated as Gaussian. This lower bound is
independent from the Higgs mass and therefore stronger than the limits
that can be derived from any other currently available
measurement. This underscores the outstanding role of the inclusive
radiative $\bar{B}$-meson decay in searches for new physics close to
the electroweak scale.

\acknowledgments{We are grateful to Andrzej Buras for suggesting the
  topic, for his careful reading of the manuscript and his valuable
  comments. Enlightning discussions with Einan Gardi concerning the
  calculation of the differential $\BXsga$ decay rate using dressed
  gluon exponentiation \cite{Andersen:2006hr} is acknowledged. We
  finally would like to thank Ignatios Antoniadis for bringing
  \cite{Antoniadis:1990ew} to our attention and for his interest in
  our article. This work has been supported in part by the Schweizer
  Nationalfonds and the National Science Foundation under Grant
  PHY-0355005.}

\end{document}